4.
\documentclass[a4paper, amsfonts, amssymb, amsmath, reprint, showkeys, 
nofootinbib, twoside, prl,superscript address]{revtex4-2}

\usepackage{graphicx}% Include figure files
\usepackage{dcolumn}% Align table columns on decimal point
\usepackage{bm}% bold math
\usepackage[mathlines]{lineno}% Enable numbering of text and display math
\usepackage[mathlines]{lineno}% Enable numbering of text and display math
\usepackage[utf8]{inputenc}
\usepackage[T1]{fontenc}
\usepackage{mathptmx}
\usepackage{xcolor}
\usepackage{float}
\usepackage{soul}
\setstcolor{red}
\begin{document}

%\preprint{AIP/123-QED}

\title[Direct laser acceleration in underdense plasmas with multi-PW lasers: a path to high-charge, GeV-class electron bunches]{Direct laser acceleration in underdense plasmas with multi-PW lasers: a path to high-charge, GeV-class electron bunches}
% Force line breaks with \\

	\author{R. Babjak}
	\email[]{robert.babjak@tecnico.ulisboa.pt}
	%\homepage[]{Your web page}
	%\thanks{}
	\affiliation{GoLP/Instituto de Plasmas e Fusão Nuclear, Instituto 
Superior Técnico, Universidade de Lisboa, Lisbon, 1049-001, Portugal}
	\affiliation{Institute of Plasma Physics, Czech Academy of 
Sciences, Za Slovankou 1782/3, 182 00 Praha 8, Czechia}

    \author{L. Willingale}
	\affiliation{Gérard Mourou Center for Ultrafast Optical Sciences, 
University of Michigan, Ann Arbor, MI 48109, United States of America}
  
    \author{A. Arefiev}
	\affiliation{University of California San Diego, La Jolla, CA 
92093, United States of America}

	\author{M. Vranic}
	\affiliation{GoLP/Instituto de Plasmas e Fusão Nuclear, Instituto 
Superior Técnico, Universidade de Lisboa, Lisbon, 1049-001, Portugal}

\date{\today}% It is always \today, today,
             %  but any date may be explicitly specified

\begin{abstract}
The direct laser acceleration (DLA) of electrons in underdense plasmas can 
provide 100s of nC of electrons accelerated to near-GeV energies using 
currently available lasers. Here we demonstrate the key role of electron 
transverse displacement in the acceleration and use it to analytically 
predict the expected maximum electron energies. The energy scaling is 
shown to be in agreement with full-scale quasi-3D particle-in-cell (PIC) 
simulations of a laser pulse propagating through a preformed guiding 
channel and can be directly used for optimizing DLA in near-future laser 
facilities. The strategy towards optimizing DLA through matched laser 
focusing is presented for a wide range of plasma densities paired with 
current and near-future laser technology. Electron energies in excess of 
10 GeV are accesible for lasers at $I\sim 10^{21}~\mathrm{W/cm^2}$.
\end{abstract}

\maketitle

Advanced acceleration schemes for electrons in plasmas allow to obtain 
multi-GeV electron bunches within a few centimeters of propagation. The 
most frequently explored scheme thus far is laser-wakefield acceleration 
(LWFA) \cite{faure2004,geddes2004,mangles2004,leemans2014}, with 7.8 GeV 
maximum energy achieved to date \cite{gonsalves2019}. One disadvantage of 
the LWFA is that it provides a relatively low number of accelerated 
electrons (10s of pC). Many applications such as x-ray and gamma-ray 
generation \cite{chen2013,stark2016,jansen2018,gunther2022}, ion 
acceleration \cite{snavely2000,wilks2001} or electron-positron pair 
creation \cite{chen2009,vranic2018,he2021,he2021_2,amaro2021,martinez2022} 
would benefit from having a high charge electron bunch, but do not require 
mono-energetic electrons. For those applications, direct laser 
acceleration \cite{pukhov1998,pukhov1999} (DLA) is a promising alternative 
to provide the electron charge on the order of 100s of nC 
\cite{hussein2021,shaw2021}. In plasmas, DLA and LWFA can act 
simultaneously \cite{shaw2017,shaw2018,king2021,lamac2021,miller2023,cohen2024} and 
in a recent experiment the obtained electron beams had energies exceeding 
10 GeV \cite{aniculaesei2022}. DLA could provide an opportunity to achieve 
an efficient energy transfer from the available laser energy into 
energetic electrons in the upcoming 10 PW-class laser facilities 
\cite{zou2015,webber2017,tanaka2020}.
According to experimental and simulation results so far, the electron 
energies are expected to vary as a function of the plasma density 
\cite{gahn1999,mangles2005,willingale2013,willingale2018,huang2017_hosing,hussein2021,rosmej2019,rosmej2020,rosmej2021,rinderknecht2021,rusby2023,vladisavlevici2023}. 
As the propagation of the laser pulse through the plasma is highly 
nonlinear and has a significant impact on the acceleration, obtaining a 
direct comparison of theory, PIC simulations and experiments is a 
challenging task. Previous theoretical models of DLA therefore use a 
simplified analytical description of the acceleration to predict the 
electron energies depending on the laser $a_0$ and the plasma density in 
ideal conditions (e.g. a plane-wave laser)  \cite{khudik2016}. 

In this Letter, we propose a new model for predicting electron energies. 
We uncover the key role of the laser width in obtaining an efficient DLA 
acceleration, which allows us to propose optimal laser focusing to achieve 
the maximum electron energy gain. Even though higher laser intensities 
generally provide higher electron energy within a shorter propagation 
distance, the relation between the laser intensity and the electron energy 
cuttoff is not linear, and we show that going for the highest possible 
laser intensity may not be the most favourable approach. Within an ion 
channel, particles perform betatron oscillations with a certain amplitude. 
An electron can experience a resonant energy gain when the frequency of 
these oscillations matches the frequency of the laser field oscillations 
at the electron location. Our analytical model predicts the maximum 
resonant amplitude for a given laser intensity. This uncovers a trade-off 
between using a highest possible laser intensity and a spotsize with a 
large enough interaction volume for optimal acceleration to occur. 
Our findings (both, analytical and from PIC simulations) show that using a 
not-so-tight focus ($\sim$ 10 $\lambda$) provides higher electron energies 
compared with when the same laser pulse is focused close to the 
diffraction limit. This comes with an added benefit that the stable laser 
guiding is easier to achieve for paraxial laser beams. We can expect to 
accelerate electrons of > 100s nC charge to multi-GeV energies within 
millimeters or centimeters of plasma.  

\begin{figure}[t]
	\includegraphics[width=.45\textwidth]{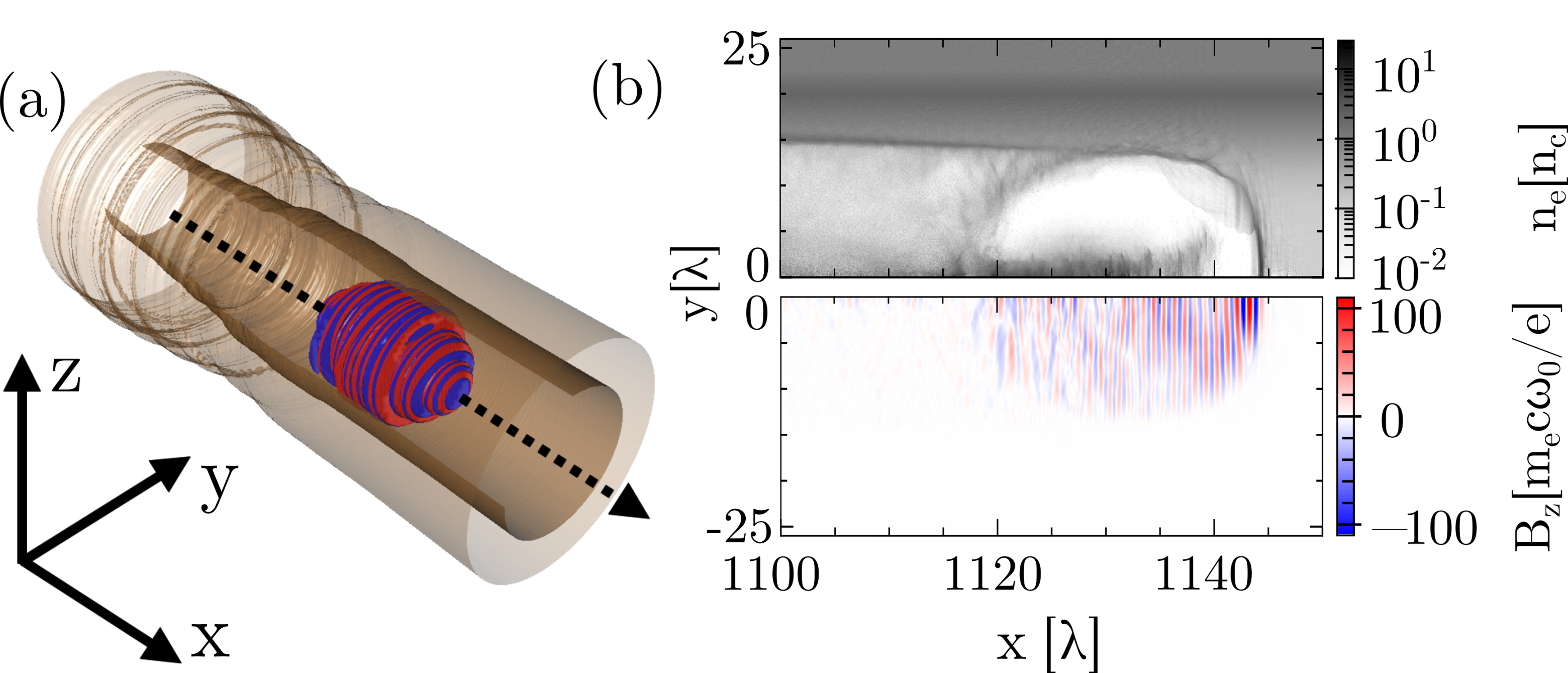}
	\caption{ (a) Schematic representation of the laser pulse 
propagating through the preformed guiding channel (b) Electron density 
during the interaction of the 10 PW laser pulse focused to $a_0=85$ 
interacting with the channel with the density at the center $n_e = 0.1n_c$ 
and the laser pulse transverse magnetic field component after $\approx$ 1 
mm of propagation. }
	\label{fig:fields}
\end{figure}

The most favorable conditions for electron acceleration using DLA are when 
a long laser pulse (100s of fs) propagates through an underdense or 
near-critical plasma. The ponderomotive force expels the electrons 
creating an ion channel, see Fig.\ref{fig:fields} (a). Quasi-static 
electromagnetic fields are generated within the ion channel, facilitating 
electron oscillations around the central axis and causing some electrons 
to be resonantly accelerated to energies exceeding the vacuum limit. The 
simplest description of DLA assumes the laser is a plane wave with a 
temporal envelope propagating within the static electric and magnetic 
fields linearly dependent on the radial distance (see Appendix) 
\cite{pukhov1999,arefiev2012,khudik2016,huang2016,huang2017}. In this 
ideal description, the equations have a conserved quantity $I = \gamma - 
p_x/m_ec + \omega_p^2y^2/4c^2$ throughout the interaction 
\cite{khudik2016}, as long as there are no losses (eg. radiation reaction 
\cite{jirka2020}) and the background plasma parameters remain constant. 
Here,  $\omega_p$ stands for the background plasma frequency, $y$ is the 
transverse displacement of the oscillating electron, $\gamma$ is the 
relativistic Lorentz factor and $p_x$ is the momentum in the direction of 
laser propagation. The integral of motion $I$ can be derived directly from 
the Hamiltonian. The plasma channel has a radial electric field and an 
azimuthal magnetic field, both acting on negatively charged particles 
co-propagating with the laser in the direction towards the channel axis 
\cite{khudik2016,arefiev2016,arefiev2016_2,vranic2018b}. This causes 
betatron oscillations with a typical frequency $\omega_{\beta}= 
\omega_p/\sqrt{2\gamma} $ \cite{pukhov2002}. Electrons simultaneously 
oscillate in the field of the laser, which is modeled as a wavepacket with 
the angular frequency $\omega_0$.
As relativistic electrons co-propagate with a laser, the resonant DLA sets 
off if the frequency of the laser field oscillations at the electron 
location $\omega'$ matches the frequency of the betatron oscillations. The 
theoretical maximum energy of the most energetic electrons is then 
determined according to their initial position and the background plasma 
density \cite{khudik2016}

\begin{equation}\label{eq:gmax}
\gamma_{\rm{max}} \simeq 2I^2\frac{\omega_0^2}{\omega_p^2}.
\end{equation}

Eq. (\ref{eq:gmax}) represents the upper limit for the energy that an 
electron with given initial conditions can reach. Typically, the electrons 
with large oscillation amplitudes achieve the highest energies for a given 
plasma density $n_p$. The exact solution for all the electrons can be 
found in \cite{khudik2016}. Once the maximum energy $\gamma_{\rm{max}}$ is 
achieved, an electron may decelerate due to the dephasing. For electrons 
starting with no transverse momentum, the maximum energy is fully 
determined by the background plasma frequency $\omega_p$ and the initial 
transverse distance from the axis $y_0$. If an electron has a finite 
initial $p_{y0}$, equivalent $y_0$ can always be found such that the 
particle has the same value of $I$ ($y_0$ then represents the first 
oscillation amplitude possibly higher than the initial distance).

Whether the electron gets accelerated depends also on the nonlinear 
amplification condition $ a_0 \omega_p/(\omega_0I^{3/2}) >\varepsilon_{cr} 
$ \cite{arefiev2014,khudik2016}, where $a_0$ is the dimensionless field 
amplitude and $\varepsilon_{cr}$ is the threshold parameter on the order 
of unity that varies depending on the electron pre-acceleration and the 
phase of the laser field where electron gets into the resonance. In the 
literature, the parameter takes values between 0.1 and 1.4, depending on 
the initial conditions and assumptions 
\cite{arefiev2014,khudik2016,arefiev2012}. We introduce a new 
interpretation of the amplification condition that connects the laser 
intensity and the plasma density with the maximum resonant transverse 
amplitude: 

\begin{equation}\label{eq:y_res}
y_{\rm{max}} 
=\frac{2c}{\omega_p}\sqrt{\left(\frac{a_0\omega_p}{\varepsilon_{cr}\omega_0}\right)^{2/3}-1}.
\end{equation}
Eq. (\ref{eq:y_res}) assumes no pre-acceleration ($\gamma-p_x \approx 1$, 
$p_{y0}=0$). It explicitly shows that a higher $a_0$ allows for the 
acceleration of electrons initially further from the channel axis when 
interacting with a plane wave. This allows higher intensity lasers to 
accelerate electrons to higher energies, even though according to Eq. 
(\ref{eq:gmax}), the maximum energy does not depend explicitly on $a_0$.  

\begin{figure}[t]
		\includegraphics[width=.45\textwidth]{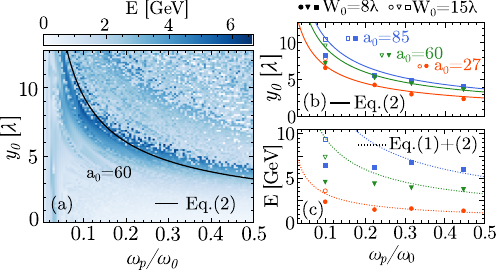}
		\caption{(a) Maximum energies achieved for test particles, 
scanning different initial values of $y_0$ and $\omega_p/\omega_0$. The 
highest energies are achieved at the maximum resonance distance 
$y_{\rm{max}}$ given by Eq. (\ref{eq:y_res}). (b) Maximum expected 
distance of resonant electrons according to Eq. (\ref{eq:y_res}) for 
different laser intensities compared with the measured maximum resonant 
distance in quasi-3D PIC simulations. (c) Maximum energies observed in 
quasi-3D PIC simulations compared with the scaling given by combining Eqs. 
(\ref{eq:gmax}) and (\ref{eq:y_res}). }
         \label{fig:resonance} 
	\end{figure}

To verify the validity of Eq. (\ref{eq:y_res}) while neglecting effects 
such as self-focusing, superluminal phase velocity \cite{yeh2021} or the 
effects arising from the finite width of focused laser pulse 
\cite{wang2019}, we first numerically integrate the equations of motion 
for an electron in an idealized ion channel with given initial conditions 
$\omega_p/\omega_0$ and $y_0$. The pulse is modeled as a transverse plane 
wave with a short longitudinal ramp-up followed by a constant field 
amplitude. The $\omega_p/\omega_0$ determines the strength of the channel 
field that is expressed as $E_y^{ch}= m_e \omega_p^2y/2e $. Without the 
loss of generality, we consider only the electric component of the channel 
field. The restitution force provided by the channel background is 
equivalent if it is composed of an azimuthal magnetic field, a radial 
electric field or the combination of both $\vec{E}$ and $\vec{B}$ 
\cite{jirka2020}. Along with the plasma frequency $\omega_p$, the second 
varied initial parameter is the transverse oscillation amplitude $y_0$. We 
tracked the particles for $\approx 6300$ laser periods (20 ps for a laser 
with $\lambda=1~\rm{\mu m}$) and extracted the maximum energy of each 
electron along its trajectory. The results for laser $a_0=60$ are shown in 
Fig. \ref{fig:resonance} (a). The black line represents Eq. 
(\ref{eq:y_res}), which predicts the maximum initial distance of electrons 
that are expected to achieve resonant motion. The resonance in test 
particle simulations (initial position $y_0$ where the maximum energy is 
obtained since $\gamma_{\rm{max}} \sim y_0^4$) is in excellent agreement 
with the parameter space expected from the analytical description in Eq. 
(\ref{eq:y_res}) if we take $\epsilon_{cr}=0.2$. Resonance can happen also 
in higher harmonics, which we observe for electrons accelerated further 
than the $y_{\rm{max}}$ but the energy gain is significantly lower for 
these cases. It is important to emphasize that there is a maximum initial 
transverse distance for resonant acceleration even during the interaction 
with a laser modeled as a transversely plane wave, which indicates that 
this is a fundamental property of the acceleration mechanism and not an 
effect of the channel or laser transverse dimension. In other words, it is 
beneficial to have a laser pulse wider than the maximum transverse 
resonant distance given by Eq. (\ref{eq:y_res}), but a further widening of 
the laser beam $W_0 \gg y_{\rm{max}}$ should not result in particle 
acceleration at a higher transverse oscillation amplitude.

The central role of $y_{\rm{max}}$ for the acceleration was also verified 
by the quasi-3D OSIRIS PIC simulations using a finite laser pulse 
propagating in a preformed plasma channel \cite{fonseca2002,davidson2015}. 
The transverse density structure of a guiding plasma channel prevents the 
laser pulse from defocusing and enables the propagation of the laser pulse 
for distances exceeding the Rayleigh length. The creation of structures 
that enable guiding of the laser pulse over several mm was previously 
studied both using PIC simulations and experimentally 
\cite{levato2020,oubrerie2022,gupta2022}. Please note that both preformed 
quasi-neutral plasma channel and an ion channel created by the laser pulse 
itself can guide the laser, but the effective interaction plasma density 
can be different for the two cases. The laser pulse with the Gaussian 
transverse profile with $W_0=8~ \rm{\mu m}$ and the 200 fs duration 
interacts with the plasma for $\approx$ 1600 laser periods (5.3 ps, 1.6 
mm) or longer if more time is needed to achieve maximum energies. The 
laser power is 1, 5, and 10 PW corresponding to the  peak $a_0$ of 27, 60, 
and 85 respectively. The simulation details are provided in the Appendix. 
The laser B field and electron density are shown in Fig. \ref{fig:fields} 
(b).

The transverse oscillation amplitude of maximum energy electrons is 
extracted from phase space diagnostic in PIC simulations. Results 
summarized in Fig. \ref{fig:resonance} (b) are compared with the expected 
values of $y_{\rm{max}}$ according to Eq. (\ref{eq:y_res}). The values 
observed in PIC simulation are in excellent agreement with our predictions 
except for the lowest density case of $n_e= 0.01 ~ n_c$. The reason for 
the discrepancy is the fact that the laser pulse was not wide enough for 
the electrons to get into the resonance at the maximum transverse 
oscillation amplitude allowed for the considered $a_0$ values (i.e. $W_0 < 
y_{\rm{max}}$). Additional simulations performed with a wider spotsize 
($W_0=15~\rm{\mu m}$) represented with the same color using empty markers 
instead of full ones agreed with Eq. (\ref{eq:y_res}). 

	\begin{figure}[t]
		\includegraphics[width=.50\textwidth]{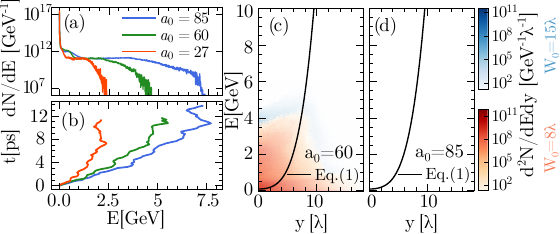}
		\caption{(a) Distribution function of electrons after the 
maximum cutoff energy is achieved. (b) Maximum energy (in the simulation) 
over time. The acceleration stops after accomplishing the theoretical 
maximum energy associated with the amplitude $y_0$. (c)-(d) Electron 
distribution in energy and instantaneous distance from the axis $y$ (not 
the same as $y_0$ as for most of electrons $y$ can be mid-oscillation). 
Each panel shows electrons accelerated by lasers with different 
intensities, but the same spotsize, where in (c) $W_0=8~\rm{\mu m}$ (d) 
$W_0=15~\rm{\mu m}$. The background plasma density is $n_e=0.01~n_c$ for 
all panels. Increasing $a_0$ resulted in electron acceleration further 
from the axis, and consequently higher maximum energies. Increasing the 
spotsize had a similar effect. }
		\label{fig:n1_energies}
	\end{figure}

We have thus shown that we are able to predict the maximum transverse 
distance from the axis of resonant electrons. This can be used to predict 
the energies of such electrons. We first calculate $y_{\rm{max}}$ 
predicted by Eq. (\ref{eq:y_res}), and then use it to obtain 
$I_{\rm{max}}\approx 1 + \omega_p^2 y_{\rm{max}}^2 / 4c^2$. Placing the 
result in Eq. (\ref{eq:gmax}) gives the maximum energy and in a compact 
form can be written as 

\begin{equation}\label{eq:gmax_resonant}
    \gamma_{\rm{max}}^{\rm{res}} = 2 
\left(\frac{a_0}{\varepsilon_{cr}}\right)^{4/3}\left(\frac{n_e}{n_c}\right)^{-1/3}
\end{equation}

The prediction is valid only if the laser pulse is wide enough and its 
propagation is stable for long enough to fully complete the acceleration. 
Using the acceleration rate over time from \cite{jirka2020}, the 
acceleration distance needed to achieve the energy 
$\gamma_{\rm{max}}^{\rm{res}}$ is $L_{acc}/\lambda = 0.78a_0^{2/3}\varepsilon_{cr}^{-5/3}(n_e/n_c)^{-2/3}$.

The prediction of the maximum energy for a given $a_0$ according to the 
above procedure is depicted by the lines in Fig. \ref{fig:resonance} (c). 
They are compared with maximum energies achieved in corresponding PIC 
simulations (shown as points). 
A higher value of $a_0$ enables particles to achieve resonance at a bigger 
$y_{\rm{max}}$, and indirectly leads to higher electron energies.

Note, that both theoretical predictions and PIC simulations neglect the 
radiation reaction. The energy is not only limited by Eq. (\ref{eq:gmax}) 
but also by the radiation reaction energy losses \cite{jirka2020}. For 
moderate laser field amplitudes used in our simulations the field of the 
ion channel is the dominant source of the radiation damping. The radiation 
limit for electron energy is proportional to $\gamma_{rr} \sim (\lambda 
a_0 \omega_0^2 /( \omega_p^2\sqrt{I}))^{2/5} $, and is more important for 
denser plasmas.

Fig. \ref{fig:n1_energies} (a) shows the energy spectrum of the electrons 
at the moment when the acceleration was complete by reaching the maximum 
allowed energy given by Eq. (\ref{eq:gmax}). The broadband energy spectrum 
is characteristic for the DLA. The charge contained within the beam can be 
defined as the number of electrons with energies higher than the vacuum 
limit given by the $\gamma_{vac} \approx a_0^2/2$. We obtain 50 nC for 
$a_0=27$, 30 nC for $a_0=60$ and 30 nC for $a_0=85$ for the channel 
density $n_e = 0.01~n_c$. For the channel density of $0.1~n_c$, the total 
charge exceeds 100 nC for the 10 PW case. The conversion efficiency for 
our simulations was in the order of 10s of percent, see Appendix. The 
maximum energy increase over time is shown in Fig. \ref{fig:n1_energies} 
(b). The maximum energy increases linearly with the propagation distance 
until saturation and the fastest energy gain is associated with the 
highest $a_0$. This is consistent with the previously derived rate for the 
energy gain over time $d\gamma / dt \sim a_0 \omega_p /\sqrt{I}$ 
\cite{jirka2020}. The energy increase stops at $\gamma_{\rm{max}}$ as 
defined by Eq. (\ref{eq:gmax_resonant}). 
This has important implications for experiments: to achieve the 
theoretical energy limit, the acceleration needs to be sustained for the 
distance $L_{acc}$.

The most energetic electrons oscillate with $y_{\rm{max}}$ around the 
channel axis, which is visible in Fig. \ref{fig:n1_energies} (c) and (d) 
that shows the two-dimensional electron histogram in energy and 
instantaneous distance from the axis. The figures demonstrate that if the 
laser pulse is not wide enough to interact with electrons at the optimal 
transverse distance ($W_0<y_{\rm{max}}$), increasing the width $W_0$ of 
the laser pulse leads to the increase in electron energy. This indicates 
that the choice of $W_0$ was not optimal for the given $a_0$. The maximum 
energy at the maximum transverse oscillation distance follows the 
theoretical prediction for both values of $W_0$. The highest energies are 
not located on-axis even though this is the region with the highest 
intensity because the electrons initially closer to the channel axis have 
a lower energy limit.  
Such a dependence of maximum energies on the oscillation amplitude 
consequently results in the forking structure of the phasespace, which is 
one of the experimental signatures of the DLA \cite{shaw2018}. Note that 
the high-energy electrons close to the $y=0$ in Fig. \ref{fig:n1_energies} 
(c) are in fact electrons with a high $y_0$ mid-oscillation. 
 
\begin{figure}[t]
		\includegraphics[width=.45\textwidth]{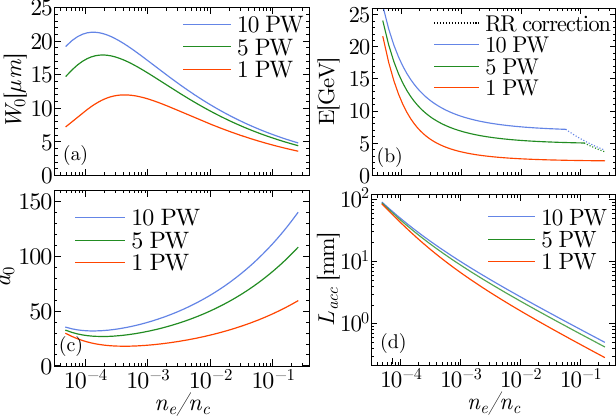}
		\caption{(a) Optimal focusing of the laser as a function 
of total power. At the given density, the maximum energy is achieved if 
the stable propagation of the laser pulse is ensured for long enough. (b) 
The value of electron energy that should be achieved at ideal focusing of 
the laser pulse according to values in (a). (c) The value of laser $a_0$ 
resulting from the optimal focusing. (d) The acceleration length needed to 
achieve the maximum predicted energies. }
		\label{fig:optimal}
	\end{figure}

Choosing the optimal width of the laser pulse is crucial for maximizing 
electron energies. 
If the laser pulse is too wide ($W_0 \gg y_{\rm{max}}$), the laser energy 
is used for the interaction with electrons beyond $y_{\rm{max}}$ that can 
not get accelerated by the most efficient first harmonic resonance. 
On the other hand, if the laser pulse is too narrow, electrons with the 
potential to achieve the highest energies ($y_0\approx y_{\rm{max}}$) can 
not get accelerated.
Therefore, the laser pulse is optimally focused when the laser waist and 
the guiding channel width are matched to the oscillation distance 
$y_{\rm{max}}$. To ensure matched conditions, an optimal focal width for a 
given laser system can be estimated analytically. We take that the 
betatron oscillations amplitude should not exceed $y \sim  W_0 / 1.2$, 
where the amplitude of the electric field is half of the maximum for a 
Gaussian pulse. This can be expressed in terms of the equation as $W_0/1.2 
= y_{\rm{max}}(\omega_p,P)$, where the field amplitude $a_0$ was replaced 
by the laser power $\rm{P}$.  
The optimal focus is then achieved if the maximum amplitude of betatron 
oscillations given by the laser width is equal to maximum resonant 
distance $y_{\rm{max}}$ expressed by Eq. (\ref{eq:y_res}). Preformed 
guiding channel prevents the laser diffraction and along with the 
self-focusing sets the upper limit on the laser waist. 

Optimal focusing can be prescribed as a function of the laser power given 
by $\rm{P~[PW]}\simeq 2.2\times  10^{-5} a_0^2 W_0^2[\rm{\mu 
m}]/\lambda^2[\rm{\mu m}]$. The optimal value of the laser waist $W_0$ for 
a given laser power P and the plasma frequency $\omega_p$ satisfies the 
following equation
\begin{equation}\label{eq:opt_w}
    W_0^2 [\lambda] = \left(\frac{\omega_0}{\omega_p} 
\frac{1.2}{\pi}\right)^2 \left[ \left( \frac{\omega_p }{\omega_0 
\varepsilon_{cr}} \sqrt{\frac{\rm{P[PW]}}{2.2\times 10^{-5}}} 
\right)^{2/3} \frac{1}{W_0^{2/3}[\lambda]} -1\right].
\end{equation}
Equation (\ref{eq:opt_w}) allows an optimal spotsize for any given laser 
system to be found, summarized in Fig. \ref{fig:optimal} (a) optical 
lasers ( $\lambda=1~\rm{\mu m}$). 
The obtained value of optimal spotsize defines  the maximum achievable 
energy shown in Fig. \ref{fig:optimal} (b). At higher density regions, 
these predictions are corrected due to radiation reaction limit. The 
corresponding peak $a_0$ in the case of optimal focus is shown in Fig. 
\ref{fig:optimal} (c). The calculations reveal the energy electrons can 
gain is higher for lower plasma densities, but they require a longer 
acceleration distance as shown in Fig. \ref{fig:optimal} (d). It was found 
out by Shaw et al. \cite{shaw2017} that the LWFA can contribute to the acceleration 
process if a significant overlap between the transverse field and trapped 
electrons is present. This happens if the dimensionless parameter 
$\omega_p \tau_{laser}/2\pi a_0 \approx 1$, where $\tau_{laser}$ is the 
laser duration. Such an overlap of both mechanisms can be expected at low 
plasma densities $n_e<0.01n_c$, potentially opening path towards 
previously unreached electron energies. 

\begin{figure}[h]
  \centering
		\includegraphics[width=.5\textwidth]{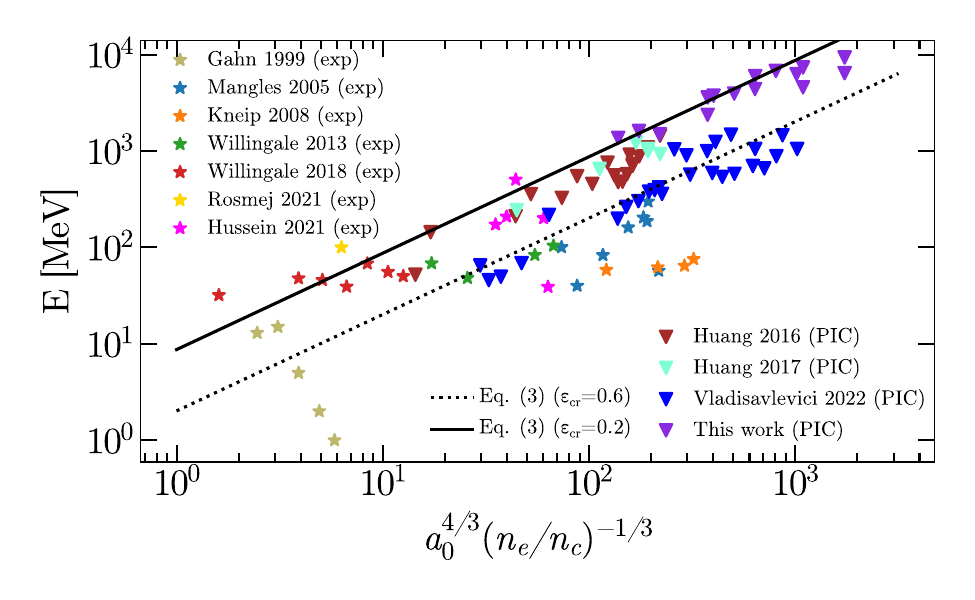}
		\caption{Comparison of cut-off energies obtained in 
various experiments and simulations  with our scaling Eq. (3) . The 
density is normalized to the critical density. Triangles depict the values 
obtained from simulations and stars depict experiments.}
    \label{fig:scaling} 

	\end{figure}

We note that the theoretical predictions presented above neglect effects 
such as strong self-focusing as well as the nonlinear effects after the 
injection of many electrons that can alter the background field structure. 
Furthermore, the value of $\varepsilon_{cr}$ influences the value of 
optimal laser width and consequent electron energies. For the solution in 
Fig. \ref{fig:optimal}, $\varepsilon_{cr} = 0.2$ was used consistently 
with the previously discussed scalings, which was also measured in PIC 
simulations to be close to 0.2 for our conditions. However, the value of 
$\varepsilon_{cr}$ can slightly vary depending on the shape of a preformed 
channel, injection from channel walls or electron pre-acceleration by the 
stochastic motion resulting from instabilities present during the 
propagation. To illustrate the validity of this choice, we have compared 
predictions of Eq. (\ref{eq:gmax_resonant}) with the cut-off energies 
extracted from several experiments and PIC simulations in Fig. 
\ref{fig:scaling}. This illustrates that the best case scenario obtained 
so far with no pre-acceleration is  $\varepsilon_{cr}=0.2$, using a wide 
range of laser intensities and plasma densities, but also that it is not 
straightforward to achieve these energies, even in simulations. To ensure 
that the maximum energy corresponding to $\varepsilon=0.2$ is reached, the 
propagation distance $L_{acc}$ needs to be ensured and the condition for 
laser focusing $W_0 \geq y_{\rm{max}}$ needs to be fulfilled to enable the 
acceleration of the most energetic electrons. 

In summary, we provide the path towards reaching highest possible energies 
of electrons in near-critical plasma and gas jets with the next generation 
of lasers while achieving the high accelerated charge and conversion 
efficiency. Due to the nature of DLA, the electron energy cutoff depends 
nontrivially on the laser intensity. Our PIC simulations in quasi-3D 
geometry demonstrate that the pulse width has a central role in the 
acceleration process, and they quantitatively agree with the predictions 
of our analytical model. We have found that for multi-pettawatt lasers, 
the best strategy is to use a wide laser focus ($\sim$10 laser 
wavelengths), low plasma density (around 0.01 $n_c$) and operate at 
moderate intensity ($a_0 \sim$ 60) as this combination of parameters 
allows beam energy cutoff $\sim$ 10 GeV. To obtain this result, it is 
necessary to ensure a stable propagation of the laser pulse, without a 
significant reduction of intensity over a few-mm distance. Laser guiding 
within a pre-formed plasma channel is a good candidate for accomplishing 
this. The simulations we presented in this work predict beams with 
superponderomotive electrons in excess of 100 nC per shot, which can be 
optimized in further studies.

\begin{acknowledgments}
The authors acknowledge fruitful discussions with Dr. B. Martinez. This 
work was supported by FCT grants CEECIND/01906/2018, 
PTDC/FIS-PLA/3800/2021 DOI: 10.54499/PTDC/FIS-PLA/3800/2021 and FCT 
UI/BD/151560/2021. We acknowledge use of the Marenostrum (Spain) and LUMI 
(Finland) supercomputers through PRACE/EuroHPC awards. This work was 
supported by the Ministry of Education, Youth and Sports of the Czech 
Republic through the e-INFRA CZ (ID:90140). L.W. and A.A. were supported 
by the Department of Energy National Nuclear Security Administration under 
Award Number DE-NA0004030.
\end{acknowledgments}

\appendix

\textit{Appendix A:Simulation parameters} - Simulations presented in this 
Letter were performed using particle-in-cell code OSIRIS in Quasi-3D 
geometry. The simulation box was $160~\mu \mathrm{m}$ long in the laser 
propagation direction and $40~\mu \mathrm{m}$ wide in the perpendicular 
direction. The grid was discretized into $\Delta x \times \Delta y = 
9600\times 864$ of cells. The timestep was $\Delta t = 
2.55\times10^{-17}\rm{s}$.
The simulations with the wider laser pulse were performed with the same 
spatial resolution using a wider simulation box (the size was increased in 
the transverse direction to $71~\mu \mathrm{m}$). For both electrons and 
ions, 32 particles per cell were used. The first two modes of the angular 
decomposition were used. The axisymmetric mode that resolves the 
self-generated channel fields and the non-axisymmetric that describes the 
linearly polarized laser field.

The transverse plasma density profile in the channel (both the density of 
electrons and ions)  was $n(y)/n_c = n_{min} + (n_{max}-n_{min}) 
(y/r_c)^5$ up to the channel radius distance $r_c=20~\mu \mathrm{m}$. The 
$n_c$ corresponds to the critical density for the laser of frequency 
$\omega_0$ and is defined as $n_c=m\varepsilon_0\omega_0^2/e^2$ where $m$ 
is the electron mass, $e$ is the electron charge and $\varepsilon_0$ is 
the vacuum permittivity. Such a channel is defined in a way that $n_{min}$ 
is the density in the center of the profile and increases up to 
$n_{max}=2~n_c$ at $y=r_c$. At $y>r_c$, the density decreases linearly to 
$~1n_c$. The "channel density" referred throughout the Letter, corresponds 
to the value of $n_{min}$ at the centre of the channel (on-axis). The 
density is uniform along the propagation direction.

The duration of the laser pulse was 200 fs with an envelope defined by the 
symmetrical polynomial function that rises as $10\tau^3-15\tau^4+6\tau^5$, 
where $\tau=\sqrt{2}/\tau_0$ and $\tau_0$ is the pulse duration in FWHM. 
The transverse laser spotsize was $W_0 = ~8 \mu m$, while transverse 
dimensionless field amplitude was defined as $a_0(y)=\exp[-y^2/W_0^2]$. 
The ratio $r_c/W_0=2.5$ was kept constant throughout the work, including 
the simulations with laser pulse $W_0=15 \mu m$. This means that the 
plasma density profile was adjusted accordingly for the simulations with 
the wider pulse.

\begin{figure}[h]
 \centering
		\includegraphics[width=.5\textwidth]{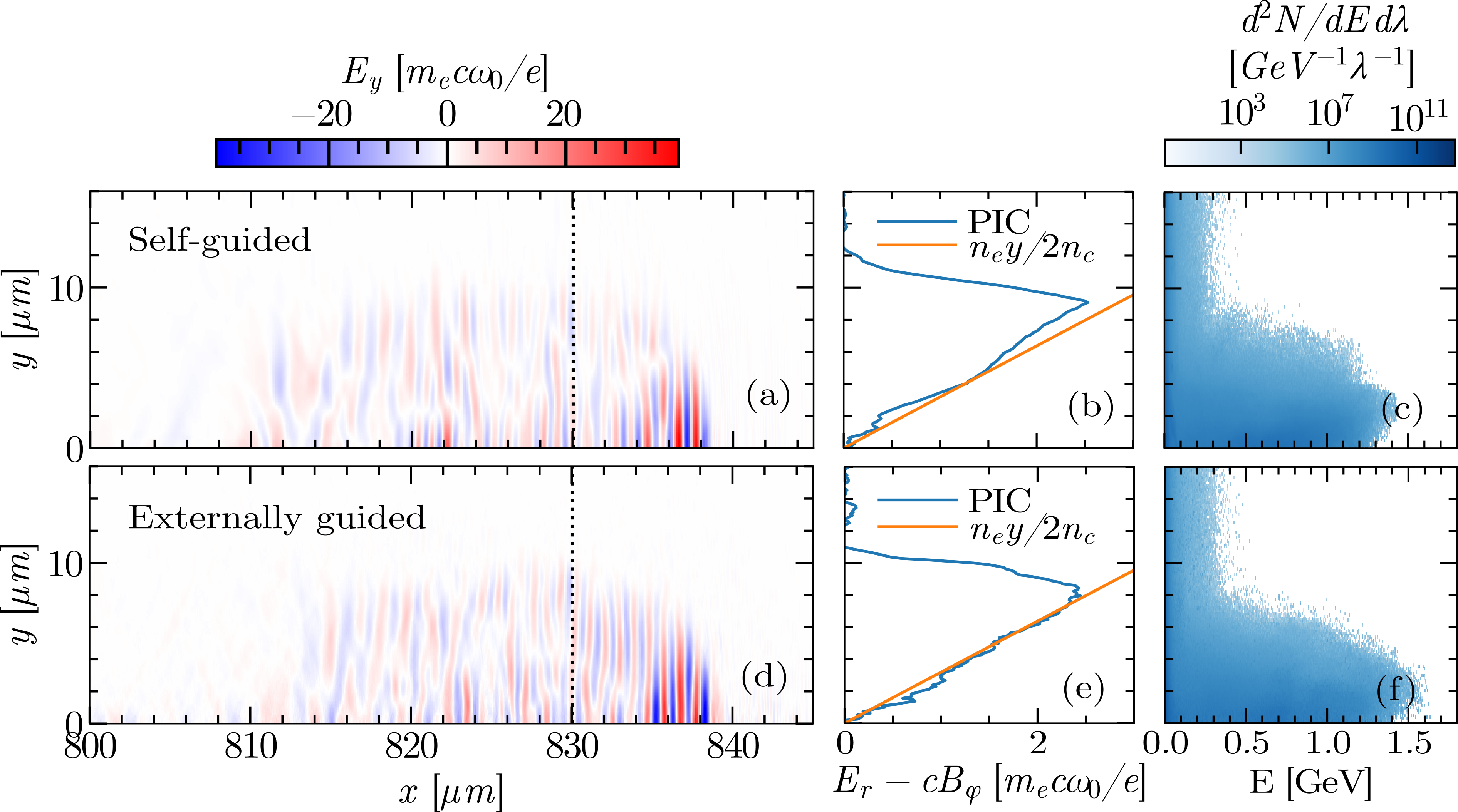}

		\caption{(a) Snapshot of the laser transverse electric 
field component at the moment, when electron maximum energies were 
achieved. (b) Line-out of the transverse ion channel focusing field at 
$x=830~\rm{\mu m}$ compared with the field used in the analytical model. 
(c) Electron energies as a function of transverse displacement at the 
moment when the maximum energies were achieved. Panels (a)-(c) depict the 
evolution in the self-guided regime (transversely constant plasma) and 
panels (d)-(f) depict the interaction in the externally guided regime with 
the preformed plasma channel. Simulation parameters are $a_0=27$, 
$\rm{W_0}=8~\rm{\mu m}$ and $n_e=0.1n_c$. In the externally guided case 
the density $n_e$ refers to the plasma density at the channel axis.}
    \label{fig:app1} 

	\end{figure}

\textit{Appendix B:The role of laser guiding by the preformed plasma 
channel} - The energy scaling law presented in this manuscript (Eq. 
\ref{eq:gmax_resonant}) is valid under the assumption of a sufficiently 
long and stable propagation distance. In our simulations, this is ensured 
by a preformed guiding structure provided by a quasi-neutral plasma 
channel. However, the laser pulse can guide itself due to the relativistic 
self-focusing for sufficiently large distances in many scenarios even 
without external guiding. In Fig. \ref{fig:app1} (a) and (d), the laser 
pulse is shown for both cases at the moment when the electrons obtain the 
maximum energy. The simulation parameters for the comparison were: 
$n_e=0.1~n_c$,~$W_0=8~\rm{\mu m}$ and $a_0=27$. After nearly a 1 mm of 
propagation, the laser shape and the intensity profile in both cases are 
comparable. The Fig. \ref{fig:app1} (b) and (e) shows the ion channel 
field structure. The transverse dependence of the quasi-static electric 
and magnetic fields responsible for the betatron oscillations are almost 
identical for the self-guided and the externally guided scenario. It is 
also worth noting the agreement of the field observed in the PIC 
simulations with the dependence used in the analytical model $E_r \sim 
n_er/2n_c$. The separation of the laser field and the channel field was 
possible due to the azimuthal decomposition of the fields 
\cite{davidson2015}. Panels (c) and (f) compare the electron 2D histogram, 
which is a function of transverse distance and the electron energy, with 
no appreciable difference noted. 
Even though the laser pulse can propagate for sufficiently long distances 
due to the self-focusing, the external guiding allows us to have better 
control over the laser spotsize during the propagation. This enables the 
laser pulse propagation with laser spotsize matched to the maximum 
transverse resonant distance, resulting in the most optimal DLA. 

\begin{figure}[h]
  \centering
		\includegraphics[width=.35\textwidth]{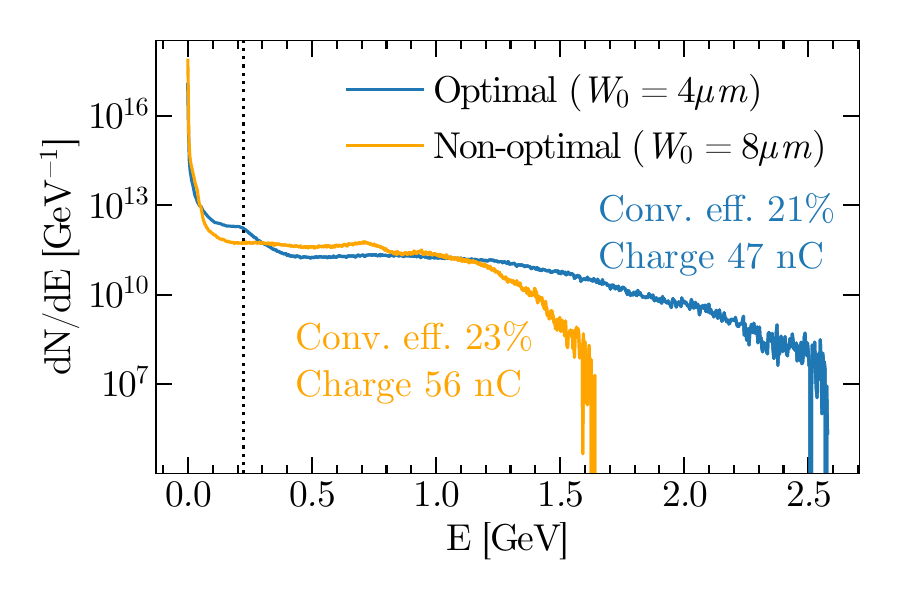}
		\caption{Distribution function of electrons accelerated by 
the 1PW laser pulse focused optimally and slightly off optimal focusing 
interacting with the plasma of density $0.1n_c$. The channel spotsize was 
controlled by the pre-formed plasma guiding channel, which resulted in 
$a_0=27$ for the case with $W_0=8~\rm{\mu m}$ and $a_0=54$ for the case 
with $W_0=4~\rm{\mu m}$. }
    \label{fig:app2} 
\end{figure}

\textit{Appendix C:The energy conversion efficiency at the optimal 
scenario} - The reader might wonder why we focused on optimizing the 
cutoff energy rather than the energy conversion efficiency. 
In fact, as shown in Fig. \ref{fig:app2}, the optimal focusing and 
non-optimal focusing examples have comparable conversion efficiency into 
hot electrons with energies over 220 MeV (21\% and 23\%).  Both 
simulations were performed for the 1 PW laser, externally guided and 
focused to different spotsizes. One simulation (non-optimal) was performed 
with the laser pulse focused to $W_0=8~\rm{\mu m}$ with $a_0=27$. In the 
other simulation (optimal according to Eq. \ref{eq:opt_w}) the laser pulse 
was focused to $W_0=4~\rm{\mu m}$ with $a_0=54$. We can see that the 
optimal focusing led to the increase in the cut-off energy from 1.6 to 2.6 
GeV, whilst the total accelerated charge is decreased by 16\% and 
conversion efficiency has changed only by a few percent.
The energy conversion efficiency is therefore not compromised by our 
optimization focused on the cut-off energy.  

It might be tempting to expect the energy conversion efficiency to be 
lower for low plasma densities, which are found optimal in terms of energy 
cutoff. This is not the case, because the DLA can be highly efficient even 
if the plasma density is decreased. For example, a simulation at an order 
of magnitude lower density $n_e=0.01~n_c$, $W_0=8~\rm{\mu m}$ and $a_0=27$ 
(1 PW) resulted in 50 nC of accelerated charge, which is comparable to the 
results with $0.1 ~n_c$ mentioned above.  The reason is the difference in 
the total interaction time. As the energy gain over time is slower at 
lower plasma densities, the propagation time (and consequently the 
propagation distance) is longer. Since the total injected charge increases 
linearly with time \cite{valenta2023}, a lower plasma density does not 
automatically decrease the total accelerated charge or the energy 
efficiency. In fact, it has been experimentally demonstrated that the 
plasma density on the order of $10^{18}-10^{19}~\rm{cm^{-3}}$ can result 
in a high total charge of 100s of nC \cite{shaw2021,hussein2021}.

\begin{figure}[h]
  \centering
		\includegraphics[width=.35\textwidth]{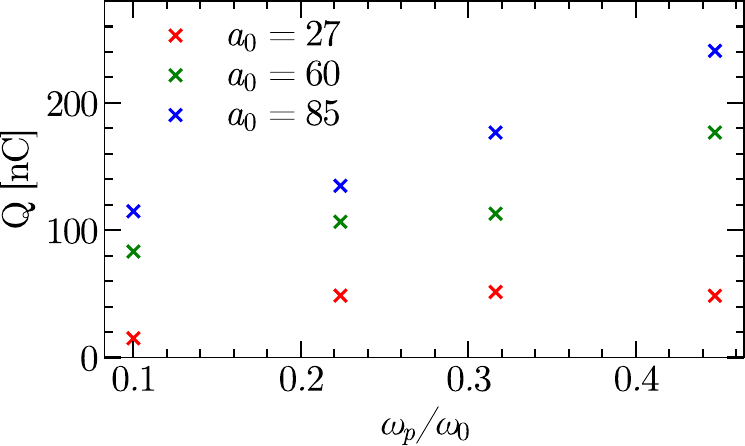}
		\caption{Total electron charge with energies exceeding 300 
MeV obtained in simulations with different plasma frequencies and laser 
intensities. The total charge was extracted at the moment when the maximum 
energy in the simulation was achieved. }
    \label{fig:app3} 
\end{figure}

\textit{Appendix D:The charge injection} - According to our scaling, 
multi-GeV energies are expected at low plasma densities. It has been 
previously demonstrated, that the accelerated charge increases with the 
propagation distance \cite{valenta2023}. As laser propagates, electrons 
are captured from the side at the laser pulse boundary or at the front of 
the pulse and the quasi-static fields of the ion channel capture those 
electrons along the axis of laser propagation. The longer laser 
propagates, the more charge can get injected. Therefore, laser propagation 
over several millimeters can compensate the slow electron injection rate 
resulting in a very high total charge.
The total injected charge for various plasma frequencies and laser 
intensities is shown in Fig. \ref{fig:app3}. The data points correspond to 
the same simulations as those in Fig. \ref{fig:resonance} (c). All 
presented simulations are with the same spotsize (of 8 microns) for 
simplicity of comparison. The total charge was taken at the moment when 
the injected charge in the simulation reached the maximum. We have 
included only electrons with energies exceeding 300 MeV. Note that the 
moment corresponding to the maximum total charge is not identical to the 
moment when electrons reach the maximum energy. In general, the charge 
injection can be happening even after the maximum energy is reached. 

Several conclusions can be drawn from Fig. \ref{fig:app3}. The total 
charge increases with the laser intensity $a_0$ and plasma density 
(represented by the plasma frequency $\omega_p$). A deeper analysis of the 
performed simulations reveals that the injection rate depends on the field 
amplitude $a_0$ and plasma frequency $\omega_p$, which then affects the 
total charge. One should expect the injection rate also to depend on the 
laser spotsize $W_0$ and the laser duration, since they influence the 
total area of the laser-dense plasma interface that determines the amount 
of available electrons. Obtaining a scaling for the total accelerated 
charge will be a subject of our future work. Nonetheless, from these 
results, one can conclude, that the direct laser acceleration by multi-PW 
lasers in underdense gas targets is capable of providing electron bunches 
with a total charge of tens or hundreds of nC.

%\nocite{*}
\bibliography{bibl}
\end{document}